\documentclass[preprint,5p]{elsarticle}
\usepackage{stmaryrd}
\usepackage{amsmath}
\usepackage{amssymb}
\usepackage{tabularx}
\usepackage{subfigure}
\biboptions{sort&compress}
\usepackage{graphicx}
\usepackage{epsfig}
\usepackage{enumitem}

\journal{Scripta Mater.}

\def\subFV{\scriptscriptstyle{FV}}
\def\subFS{\scriptscriptstyle{FS}}
\def\subVS{\scriptscriptstyle{VS}}

\begin{document}

\begin{frontmatter}

\title{Solid-State Dewetting and Island Morphologies in Strongly Anisotropic Materials}
\author[1,2]{Wei Jiang\corref{5}}
\address[1]{School of Mathematics and Statistics, Wuhan University, Wuhan, 430072, China}
\address[2]{Computational Science Hubei Key Laboratory, Wuhan University, Wuhan, 430072, China}
\ead{jiangwei1007@whu.edu.cn}
\cortext[5]{Corresponding author.}

\author[3]{Yan Wang}
\address[3]{Department of Mathematics, National University of
Singapore, Singapore, 119076}

\author[3]{Quan Zhao}

\author[4]{David J. Srolovitz}
\address[4]{Departments of Materials Science and Engineering {\rm \&}
Mechanical Engineering and Applied Mechanics, University of Pennsylvania, Philadelphia, PA 19104, USA}

\author[3]{Weizhu Bao}
\ead{matbaowz@nus.edu.sg}


\begin{abstract}
We propose a sharp-interface continuum model based on a thermodynamic variational approach to investigate the strong anisotropic effect on solid-state dewetting including contact line dynamics. For sufficiently strong surface energy anisotropy, we show that multiple equilibrium shapes may appear that can not be described by the widely employed Winterbottom construction, i.e., the modified Wulff construction for an island on a substrate. We repair the Winterbottom construction to include multiple equilibrium shapes and employ our evolution model to demonstrate that all such shapes are dynamically accessible.
\end{abstract}




\end{frontmatter}


Solid-state dewetting is a ubiquitous phenomenon in thin film technology~\cite{Thompson12,Jiran,Ye,Leroy12,Rabkin14} which can either be deleterious, destabilizing a thin film structure, or advantageous, leading to the controlled formation of an array of nanoscale particles, e.g., used in sensor devices~\cite{Mizsei93} and as catalysts for  the growth of carbon or semiconductor nanowires~\cite{Randolph07,Schmidt09}. Recently, solid-state dewetting has been attracting increased attention both  because of  interest in the underlying pattern formation physics and its potential application as an economical approach to obtain nanostructured surfaces and nanodevices~\cite{Srolovitz86,Wong00,Du10,Dornel06,Pierre09,Pierre11,Carter95,Jiang12,Zucker13,Jiang15}.

The dewetting of thin solid films deposited on substrates is similar to the dewetting of liquid films~\cite{Gennes85,Ren11}. However,  mass transport during solid-state dewetting is usually dominated by surface diffusion rather than fluid dynamics. Solid-state dewetting can be modeled as interface-tracking problem where  morphology evolution is governed by surface diffusion and contact line migration~\cite{Jiang12,Jiang15}. In  early studies, a number of simplifying assumptions were made in order to keep the analysis tractable. For example, under the assumption that all interface energies are isotropic, Srolovitz and Safran~\cite{Srolovitz86} proposed a sharp-interface  model to analyze  hole growth; based on the above model, Wong {\it et al.}~\cite{Wong00,Du10} designed a ``marker particle'' numerical scheme to study the two-dimensional retraction of an island and a perturbed cylindrical wire on a substrate. Recently, we~\cite{Jiang12} solved a similar problem using a phase field approach that naturally captures the topological events that occur during evolution and is applicable in any number of dimensions.

However, many experiments have demonstrated that the morphology evolution that occurs during thin solid film dewetting is strongly affected by crystalline anisotropy~\cite{Ye}. Recent approaches that incorporate crystalline anisotropy have included a discrete model~\cite{Dornel06}, a kinetic Monte Carlo method~\cite{Pierre09,Pierre11} and the crystalline method~\cite{Carter95,Zucker13}. The main drawback of these approaches is that the evolution does not account for the full anisotropic free energy of the system or do not represent a completely mathematical description. To overcome these shortcomings, we \cite{Jiang15} proposed a continuum model for simulating morphology evolution during solid-state dewetting for weakly anisotropic surface energies. But it is not straightforward to extend this approach to the strongly
anisotropic case, and the major difficulty comes from how to understand the thermodynamic variation including contact
line migration. In this letter, we extend this dynamical evolution continuum model to include the common case where the anisotropy is strong and its influence on solid-state dewetting morphologies is pronounced.

We note at the outset, that although we apply this dynamical evolution model to the simulation of morphology evolution during the solid-state dewetting of thin films, it also naturally provides a much more general solution to the problem of how  to determine the equilibrium shape of a crystalline island on a substrate than is currently available. This is a problem of long-standing in the materials science and applied mathematics communities;
receiving important attentions from many researchers over more than one hundred years~\cite{Gibbs1878,Wulff1901,Kaishew50,Mullins62,Winterbottom67,Taylor92,Korzec14}.
This problem can be stated as follows: determine the island shape that minimizes the total interface energy,
\begin{equation}
\mathop{\textbf{min}}\limits_{\Omega}\;W_1=\int_{\Gamma} \gamma(\theta)\;d\Gamma +
\underbrace{(\gamma_{\scriptscriptstyle {FS}}-
\gamma_{\scriptscriptstyle {VS}})(x_c^r-x_c^l)}_{{\textbf {Substrate\;Energy}}}, 
\label{eqn:crystal}
 \end{equation}
where $\Omega$ denotes the region occupied by the island, the volume of the island is conserved, i.e., $|\Omega|={\text{constant}}$, $\Gamma$ represents the film (or island)/vapor interface, and the right and left contact points are $x_c^r$ and $x_c^l$ (these are points/lines where the vapor, film and substrate coexist), and $\gamma_{\subFV}$, $\gamma_{\subFS}$ and $\gamma_{\subVS}$ are, respectively,  the surface energy densities of the film/vapor, film/substrate and vapor/substrate interfaces. We assume that the film/vapor interface energy (density) $\gamma_{\subFV}$ is a function only of the interface  normal, i.e., $\gamma_{\subFV}=\gamma(\theta)$, $\theta \in [-\pi,\pi]$ represents the local orientation of the outer normal to the film/vapor interface, and $\gamma_{\subFS}$ and $\gamma_{\subVS}$
are independent constants. The solution to problem~\eqref{eqn:crystal} yields an equilibrium shape
with minimal interface/surface energy of prescribed area (or volume).

As is well known, if the island is free-standing (i.e., not in contact with the substrate), the equilibrium shape is given by the classical Wulff construction~\cite{Gibbs1878,Wulff1901,Mullins62,Taylor92}. If, on the other hand, the island is in contact with a flat, rigid substrate, the equilibrium shape is classically described using the Winterbottom construction~\cite{Kaishew50,Winterbottom67}.
However, when the surface energy anisotropy is strong, the Wulff envelope may include  ``ears'';  cutting off  the ``ears'' gives the equilibrium shape~\cite{Mullins62,Eggleston01}. In the case of an island on a substrate, however, the existence of  ``ears''  in the Wulff envelope can give rise to  multiple stable (or metastable) shapes. As we demonstrate below, the existence of such additional states has a profound effect on morphology evolution; giving rise to stable morphologies never seen on the basis of the widely accepted and applied Winterbottom construction. Incorporation of such non-Winterbottom effects is essential to describing observed island morphologies that arise during kinetic phenomena such as the solid-state dewetting process discussed here.

We first derive the dynamical evolution model directly from the free energy, including the effect of strong interface energy anisotropy. The total free energy of the system for solid-state dewetting problems under strongly anisotropic conditions can be written in two parts:
 $W=W_1+W_2,$
where the first term $W_1$ was defined in Eq.~\eqref{eqn:crystal}, above (also see \cite{Jiang15}).
When the surface energy anisotropy is sufficiently large, the surface diffusion evolution equations become ill-posed. To address this issue, we add a regularization term $W_2$ (i.e., a Willmore energy regularization) into the system~\cite{Carlo92,Gurtin02,LiBo09,Torabi10}:
\begin{equation}
W_2=\frac{\varepsilon^2}{2} \int_{\Gamma} \kappa^2\;d\Gamma,
\label{energy:reg}
\end{equation}
where $\varepsilon$ is a small regularization parameter and $\kappa$ denotes the curvature of the film/vapor interface, $\Gamma$.

We calculate the first variation of the energy functional $W$ with respect to the interface shape $\Gamma$ and the left and right moving contact points, $x_c^l$ and $x_c^r$~\cite{Note1}.
Then, following a  procedure similar to that in the weakly anisotropic case~\cite{Jiang15},
we find that the two-dimensional solid-state dewetting of a thin film with strongly anisotropic surface energies on a flat solid substrate can be described in the following dimensionless form in a sharp-interface model (see Supplemental Material for more details):
\begin{align}
    \frac{\partial{\mathbf{X}}}{\partial t}&=V_n \mathbf{n} = \frac{\partial^2 \mu}{\partial s^2} \mathbf{n},
    \label{strongly1}\\
    &\mu=\Big(\gamma(\theta)+\gamma\,''(\theta)\Big)\kappa-\varepsilon^2\Big(\frac{\partial^2\kappa}{\partial s^2}+\frac{\kappa^3}{2}\Big),
   \label{strongly2}
\end{align}
where $\Gamma=\mathbf{X}(s,t)=(x(s,t),y(s,t))$ represents the moving film/vapor interface, $s$ is the arc length or distance along the interface and $t$ is the time, $V_n$ is the velocity of the interface in the direction of its outward normal, $\mathbf{n}$ is the interface outer unit normal direction and $\mu$ denotes the chemical potential.
Note that all lengths and interface energies are scaled by two constants $R_0$ and $\gamma_0$,  chosen as described below. The governing equations~\eqref{strongly1}-\eqref{strongly2} are subject to the following dimensionless boundary conditions:
\setitemize{fullwidth,itemindent=\parindent,listparindent=\parindent,itemsep=0ex,partopsep=0pt,parsep=0ex}
\begin{itemize}
\item[(I)] Contact point condition ({{BC1}})
\begin{equation}
y(0,t)=0, \qquad y(L,t)=0,
\label{eqn:stronglyBC1}
\end{equation}
where $L=L(t)$ denotes the total length of the interface at time $t$, and therefore we can use
$s=0$ and $s=L$ to represent the left and right contact points ($x_c^l$ and $x_c^r$).

\item[(II)] Relaxed contact angle condition ({{BC2}})
\begin{equation}
\frac{d x_c^l}{d t}=\eta f_\varepsilon(\theta_d^l), \qquad \frac{d x_c^r}{d t}
=-\eta f_\varepsilon(\theta_d^r),
\label{eqn:stronglyBC2}
\end{equation}
where $\theta_d^l$ (or $\theta_d^r$) is the (dynamical) contact angle at the left (or right)
contact point, $\eta$ represents the contact line mobility,
$f_\varepsilon(\theta)=\gamma(\theta)\cos\theta-\gamma\,'(\theta)
\sin\theta-\sigma-\varepsilon^2\frac{\partial\kappa}{\partial s}\sin\theta$,
and the material parameter $\sigma=(\gamma_{\scriptscriptstyle {VS}}-\gamma_{\scriptscriptstyle {FS}})/\gamma_0$.

\item[(III)] Zero-mass flux condition ({{BC3}})
\begin{equation}
\frac{\partial \mu}{\partial s}(0,t)=0, \qquad \frac{\partial \mu}{\partial s}(L,t)=0.
\label{eqn:stronglyBC3}
\end{equation}

\item[(IV)]Zero-curvature condition ({{BC4}})
\begin{equation}
\kappa(0,t)=0, \qquad \kappa(L,t)=0.
\label{eqn:stronglyBC4}
\end{equation}
\end{itemize}
Because these dynamical evolution PDEs are
sixth-order (fourth-order  for  weak anisotropy~\cite{Jiang15}),
to make the system well-posed, we introduced an additional boundary condition ({{BC4}}), which rigorously comes from
the variation of the total energy functional~\cite{Note2}.
The total free energy of the system described by Eqs.~\eqref{strongly1}-\eqref{eqn:stronglyBC4} can be shown to decrease monotonically at all times and that the total mass of the solid film on top of the substrate is conserved
during the evolution.
\begin{figure*}[!tp]
\centering
\includegraphics[width=15cm]{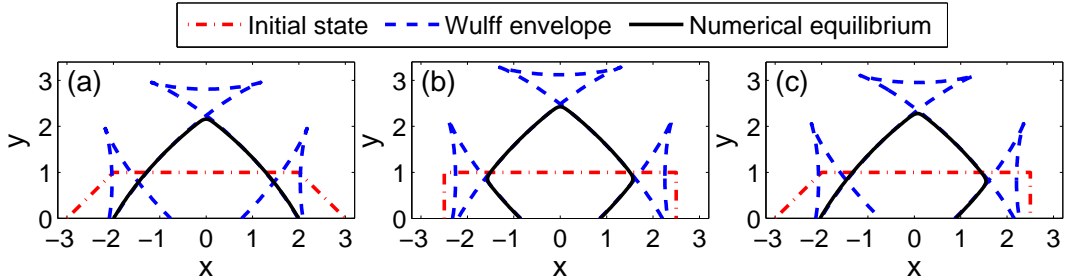}
\caption{The equilibrium shapes of thin films with different initial island shapes (shown by dash-dot red lines)
under the same parameters: $m=4, \beta=0.3, \sigma=-0.5$, where the solid black lines show the different
numerical equilibrium shapes, and the dashed blue lines represent the Wulff envelope truncated by
the flat substrate. We note here that the equilibrium shape may include some parts of the ``ears'' seen in
the Wulff envelope when the anisotropy of surface energy is strong.
}
\label{fig:winterbottom}
\end{figure*}

We solve the governing equations~\eqref{strongly1}-\eqref{eqn:stronglyBC4} by using a parametric semi-implicit mixed finite element scheme~\cite{Bao15}. Compared to traditional explicit finite difference approaches (e.g., marker particle methods), the proposed finite element method allows for larger time steps while satisfying numerical stability requirements~\cite{Barrett}.
We set the initial film thickness to unity (i.e., we choose $R_0$ as the initial film thickness) and assume a dimensionless anisotropic surface energy of the form:
\begin{equation}
  \gamma(\theta) = 1+\beta\cos [m(\theta+\phi)],
  \label{eqn:sfenergy}
\end{equation}
where $\beta$ controls the magnitude of the anisotropy, $m$ is the  rotational symmetry order
and $\phi$ represents a phase shift angle describing a rotation of the crystallographic
axes of the island with respect to the substrate plane ($\phi$ is set to
zero except where noted). It should be pointed out that although we assume that the surface energy
is smooth, for the non-smooth or ``cusped'' surface energy, we can deal with the problem by smoothing the
surface energy with small parameters.

We now turn to the issue: how does strong anisotropy affect solid film dewetting morphologies - especially, the stable island shapes produced by dewetting? In the proposed model,
we find that if the small regularization parameter $\varepsilon$ goes to zero, the equilibrium contact angles $\theta_a$  satisfies the anisotropic Young equation~\cite{Jiang15,Min06}:
\begin{equation}\label{eqn:forcebalance}
\lim_{\varepsilon\to 0} f_{\varepsilon}(\theta)=f(\theta)= \gamma(\theta) \cos\theta - \gamma\,'(\theta) \sin\theta - \sigma = 0.
\end{equation}
Eq.~\eqref{eqn:forcebalance} may have multiple roots in $\theta\in [0,\pi]$ only when
there exist orientations for which $\gamma(\theta)+\gamma\;''(\theta)<0$,
i.e., in strongly anisotropic cases.

We perform a series of dynamical evolution numerical simulations for different initial island shapes for strongly anisotropic surface energies at fixed island volumes. Several examples are shown in Fig.~\ref{fig:winterbottom} for $m=4, \beta=0.3, \sigma=-0.5$. We clearly see that depending on the initial island shape, three different stable shapes evolve (shown by the solid black lines).

All three stable shapes can be  predicted by generalizing the Winterbottom construction, i.e., using the flat substrate to truncate the Wulff envelope (shown by the dashed blue lines) in Fig.~\ref{fig:winterbottom} where the truncations may include  parts of the ``ears''. The equilibrium Winterbottom (global minimum energy) shape is that shown
in Fig.~\ref{fig:winterbottom}(b);  the other two stable shapes
(Figs.~\ref{fig:winterbottom}(a) and (c)) correspond to local minimum energy (metastable) shapes.
These local minima can be understood in terms of the multiple roots of the anisotropic Young equation; only these roots correspond to candidate stable contact angles. In this example, the anisotropic Young equation, Eq.~\eqref{eqn:forcebalance}, has three distinct roots in $[0,\pi]$ -- corresponding to two stable and one unstable equilibrium (``stable'' equilibrium contact angles could, in principle, be obtained from the dynamical evolution).

\begin{figure}[htp]
\includegraphics[width=8.6cm]{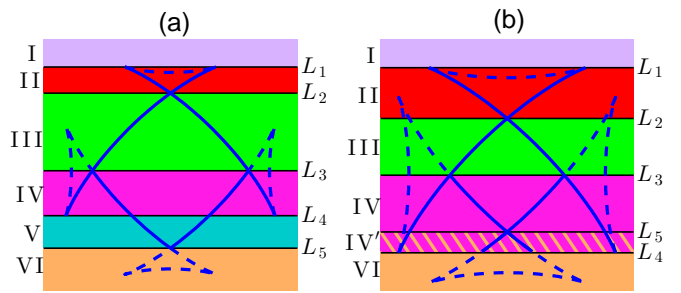}
\caption{Classification of the wetting/dewetting into six different cases for a four-fold crystalline thin film: (a)~$\beta$ is not very large, (b)~$\beta$ is very large (i.e., large ``ears''). Here, the blue curves represent the Wulff envelope, and the dashed blue curves correspond to unstable solutions of the anisotropic Young equation, Eq.~\eqref{eqn:forcebalance}.}
\label{fig:classification}
\end{figure}

\begin{figure}[htp]
\centering
\includegraphics[width=8.0cm,height=11.0cm]{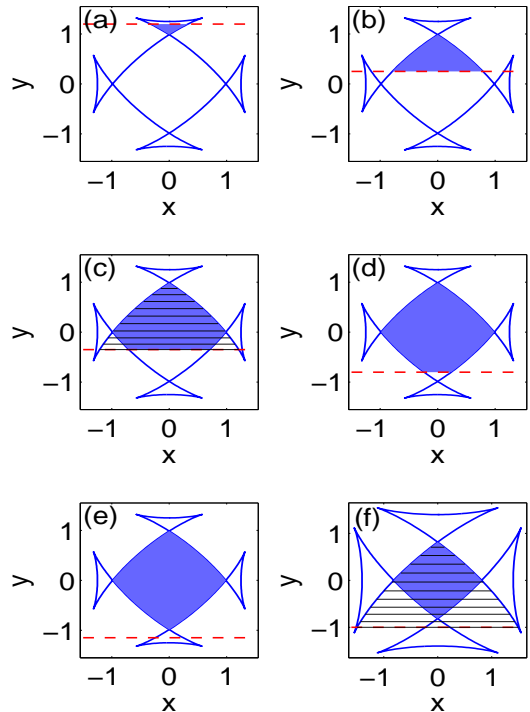}
\caption{A schematic illustration of the different cases of wetting/dewetting for a four-fold crystalline film as defined in Fig.~\ref{fig:classification}:
(a) Case II, (b) Case III, (c) Case IV, (d) Case V, (e) Case VI, and (f) Case IV$^\prime$. The dashed red lines are  the flat substrates lines. The shaded blue region corresponds to the equilibrium Winterbottom island shapes and the horizontal dashed shaded regions represent other stable island shapes.}
\label{fig:wulff_multi}
\end{figure}

Although non-Winterbottom shapes have been observed in the experiments~\cite{Rabkin11a,Rabkin11b,Rabkin15}, there exists very few theoretical literature which can offer a good explanation and prediction. Based on a wide range of numerical results, we have developed a method to repair the classical Winterbottom construction. For any given anisotropic surface energy (i.e., $m$ and $\beta$ here), we first construct the Wulff envelope. Then, following the Winterbottom construction procedure for a flat substrate, we truncate the Wulff envelope at a height $y=\sigma$ (see Fig.~\ref{fig:wulff_multi}) to obtain the possible stable shapes. For simplicity, we describe the procedure for four-fold crystalline anisotropy, as illustrated in Fig.~\ref{fig:classification} and \ref{fig:wulff_multi} for several $(\sigma, \beta)$ combinations. Referring to Fig.~\ref{fig:classification}, we identify six different types of strongly anisotropic ($\beta>1/15$) behaviors:
\begin{itemize}
\item Case I: \emph{Complete Wetting}. The ``substrate line'' $L_1$ falls above the Wulff envelope such that $f(\theta)$ is always less than zero. In this case, for any initial island shape,  the contact points will  move outward and there is no stable shape; the island tends to cover the substrate.

\item Case II: \emph{Partial Wetting}, $\theta_a \in (0,\frac{\pi}{2})$. The equilibrium shape is found by flipping over the part of the Wulff envelope truncated by the substrate line that lies between $L_1$ and Line $L_2$, as indicated by the blue shaded region in Fig.~\ref{fig:wulff_multi}(a) for the red dashed substrate line.

\item Case III: \emph{Partial Wetting}, $\theta_a\in(0, \frac{\pi}{2})$. The equilibrium shape can be directly obtained from the section of the Wulff shape delimited by the substrate line between $L_2$ and Line $L_3$ in Fig.~\ref{fig:classification}, as shown by the blue shaded region in Fig.~\ref{fig:wulff_multi}(b).

\item Case IV (or IV$^\prime$): \emph{Multiple Equilibrium Shapes}. In these cases, multiple stable shapes exist that can be determined by proper truncation of the Wulff envelope (shown in Fig.~\ref{fig:wulff_multi}(c)).
    In this case, there are two ``stable'' equilibrium contact angles $\theta_a \in (0,\pi)$ which yield three possible equilibrium shapes. Referring to Fig.~\ref{fig:wulff_multi}(c), the stable shapes are (i) the blue shaded region (i.e., the equilibrium Winterbottom shape), (ii) the  striped region, (iii) the left side of the island corresponds to the striped  and the right side to the blue regions, and (iv) the right side of the island corresponds to the striped  and the left side to the blue regions (the mirror of case (iii)).
    The dynamical evolution for this case with different initial conditions was shown in Fig.~\ref{fig:winterbottom},
    realizing three of these cases.

\item Case V: \emph{Partial Wetting}, $\theta_a \in (\frac{\pi}{2}, \pi)$.
The equilibrium shape is obtained from the section of the Wulff shape delimited by the substrate line between Lines $L_4$ and  $L_5$ in Fig.~\ref{fig:classification}, as shown by the blue shading in Fig.~\ref{fig:wulff_multi}(d).

\item Case VI: \emph{Complete Dewetting}. This case corresponds to complete dewetting (shown in Fig.~\ref{fig:wulff_multi}(e)).

\end{itemize}
Note here that when $\beta$ becomes very large, the boundary line $L_4$ may fall below Line $L_5$ (shown in Fig.~\ref{fig:classification}(b)), and if the substrate line lies between $L_5$ and $L_4$, it will produce
Case IV$^\prime$. Compared to Case IV for multiple equilibrium shapes, Case IV$^\prime$ differs only
in that its equilibrium Winterbottom shape corresponds to complete dewetting
(shown in Fig.~\ref{fig:wulff_multi}(f)).

Additional numerical examples are presented in the Supplemental Material.
Expressions for the boundary lines from $L_1$ to $L_5$ are given in the Supplementary Materials (either explicitly or implicitly) for all $(\sigma,\beta)$ parameter pairs. This completely determines which shapes may be found in all cases. Although we have focused on the case of a crystal with a four-fold rotational symmetry ($m=4$) above, similar discussion can be also performed for other values of $m$.

\begin{figure}[!tp]
\centering
\includegraphics[width=8.5cm]{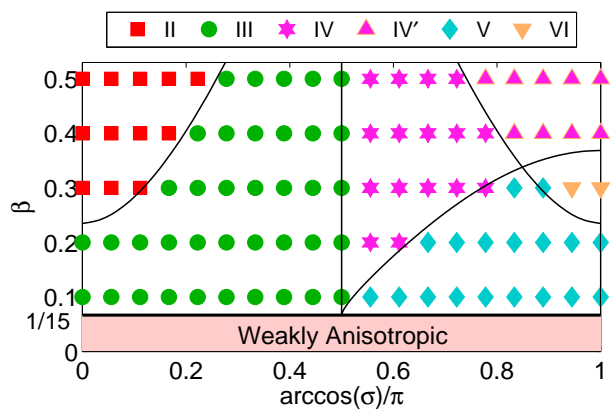}
\caption{Phase diagram between the parameter pairs and the cases of wetting/dewetting.
The colored symbols respectively denote the different cases of wetting/dewetting,
which are obtained from our numerical simulations under different parameter pairs
$(\sigma,\beta)$. The solid black lines, which are calculated
from our theoretical predictions presented in Fig.~\ref{fig:classification},
represent the different boundary lines between the different cases.}
\label{fig:winterbottom_phase}
\end{figure}

To validate the predictions shown above, we have performed ample numerical simulations to
confirm that the parameter pairs $(\sigma,\beta)$ will fall into the expected stable island shape Cases. These results are summarized in the phase diagram of
Fig.~\ref{fig:winterbottom_phase}. This diagram confirms that the numerical results
are  consistent with the theoretical predictions.


In this letter, we presented a sharp-interface continuum model for solid-state dewetting which can include the strong anisotropy effect. Based on the model, we find both the equilibrium Winterbottom and metastable
island shapes, which is unlike in the classical (Winterbottom) prediction about the shape of anisotropic islands
on a substrate. Then we proposed a theory to repair Winterbottom prediction. By starting with different film
shapes (initial conditions), our numerical simulations demonstrate that dewetting can lead to either the equilibrium Winterbottom shape or any of the metastable shapes found in our theory. While the presented results are for two dimensions, our approach can be directly generalized to three dimensions, but the main challenge is how to
design efficient and accurate numerical algorithms to simulating moving open surfaces coupled with contact line migration
in three dimensions. We believe that our approach opens the door to quantitatively simulating common cases for solid-state dewetting at large scale, and tailoring island shapes and hence subsequent material properties.

\section*{Acknowledgements}
This work was supported by the National Natural Science Foundation of China Nos. 11401446 and
11571354 (W.J.), the Fundamental Research Funds for the Central Universities (Wuhan University, Grant No. 2042014kf0014) (W.J.), the Singapore A*STAR SERC PSF-grant 1321202
067 (W.B.) and the Center for the Computational Design of Functional Layered Materials, an Energy Frontier Research Center funded by the U.S. Department of Energy (DOE), Office of Science, Basic Energy Sciences (BES) under Award \# DE-SC0012575 (D.J.S.). Part of the work was performed in 2015
when the authors were visiting Beijing Computational Science Research Center.


\end{document}